\documentclass[aps,twocolumn,nofootinbib]{revtex4}

\usepackage{amsmath,amssymb}
\usepackage{pict2e}
\usepackage{graphicx}
\usepackage{url}

\newcommand{\<}{\langle}
\renewcommand{\>}{\rangle}

\def\be#1\ee{\begin{equation}#1\end{equation}}
\def\ba#1\ea{\begin{align}#1\end{align}}
\def\bas#1\eas{\begin{align*}#1\end{align*}}
\newcommand{\nn}{\nonumber\\}

\newcommand{\eq}[1]{(\ref{eq:#1})}
\newcommand{\fig}[1]{Fig.~\ref{fig:#1}}
\newcommand{\sect}[1]{Sec.~\ref{sec:#1}}

\newcommand{\ii}{\mathrm{i}}
\newcommand{\jbar}{{\bar \jmath}}
\renewcommand{\d}{\mathrm{d}}
\newcommand{\dbar}{{\mathchar'26\mkern-12mu {\mathrm d}}}
\newcommand{\intk}[1]{\int_{-\pi}^0 \!\! #1 \, \dbar{k}}
\newcommand{\poly}{\mathrm{poly}}
\newcommand{\norm}[1]{\| #1 \|}
\newcommand{\Norm}[1]{\left\| #1 \right\|}
\newcommand{\floor}[1]{\lfloor #1 \rfloor}

\newcommand{\incoming}{{\rm sc}^\rightarrow}
\newcommand{\outgoing}{{\rm sc}^\leftarrow}
\newcommand{\bound}{{\rm bd}}

\newcommand{\rmin}{{\rm in}}
\newcommand{\rmout}{{\rm out}}

\newcommand{\Z}{{\mathbb{Z}}}
\newcommand{\SU}{{\mathrm{SU}}}

\newcommand{\cR}{\mathcal{R}}
\newcommand{\cT}{\mathcal{T}}

\newcommand{\alab}{\mathrm{(a)}}
\newcommand{\blab}{\mathrm{(b)}}
\newcommand{\clab}{\mathrm{(c)}}
\newcommand{\dlab}{\mathrm{(d)}}
\newcommand{\elab}{\mathrm{(e)}}

\newcommand{\rmb}{\mathrm{b}}
\newcommand{\rmc}{\mathrm{c}}
\newcommand{\rmd}{\mathrm{d}}

\begin{document}

\title{Universal computation by quantum walk}

\author{Andrew M. Childs}
\email[]{amchilds@uwaterloo.ca}

\affiliation{Department of Combinatorics \& Optimization
             and Institute for Quantum Computing,
             University of Waterloo
            }


\begin{abstract}
In some of the earliest work on quantum mechanical computers, Feynman showed how to implement universal quantum computation by the dynamics of a time-independent Hamiltonian.  I show that this remains possible even if the Hamiltonian is restricted to be a sparse matrix with all entries equal to $0$ or $1$, i.e., the adjacency matrix of a low-degree graph.  Thus quantum walk can be regarded as a universal computational primitive, with any desired quantum computation encoded entirely in some underlying graph.  The main idea of the construction is to implement quantum gates by scattering processes.
\end{abstract}

\maketitle

\section{Introduction}
\label{sec:intro}

Quantum computers are hypothetical devices that represent and process information according to the principles of quantum mechanics.  If built, they could solve certain problems, such as factoring integers \cite{Sho97}, dramatically faster than classical computers.

While the earliest algorithms for quantum computers were based on Fourier sampling, the concept of \emph{quantum walk} \cite{FG98,Wat01a} subsequently led to a new class of quantum algorithms.  A quantum walk is a quantum mechanical analog of a classical random walk.  By exploiting interference effects, quantum walks can outperform random walks at certain computational tasks: there are black-box problems for which quantum walk provides an exponential speedup over classical computation \cite{CCDFGS03,CSV07}, and many quantum walk algorithms achieve polynomial speedup over classical computation for problems of practical interest \cite{SKW02,CG03,Amb07,MSS05,AKR04,CG04,BS06,MN05,FGG07,ACRSZ07,RS07}.

There are several ways to define a quantum analog of a random walk, but perhaps the simplest is the continuous-time quantum walk generated by the adjacency matrix $A$ of some graph \cite{FG98}.  This quantum walk takes place on a Hilbert space spanned by orthonormal basis states corresponding to vertices of the graph, and the evolution of the walk for time $t$ is described by the unitary operator $e^{-\ii A t}$.  In other words, the quantum walk is simply the Schr{\"o}dinger dynamics of a particle allowed to hop between adjacent vertices of the graph.

The goal of this article is to explore the power of quantum walk as a general model of computation.  In particular, I show that even a rather restricted kind of quantum walk is \emph{universal} for quantum computation, meaning that any problem that can be solved by a general-purpose quantum computer can also be solved by a quantum walk.
(This parallels the situation for adiabatic evolution, which is also universal \cite{ADKLLR04,KR03,KKR04,AGIK07}.)
There are at least two motivations for demonstrating this universality.  First, it shows that quantum walk is computationally powerful, since in principle, any quantum algorithm can be recast as a quantum walk algorithm.
Indeed, these two models are equivalent, since the quantum walk on any sufficiently sparse graph can be efficiently simulated by a universal quantum computer \cite{AT03,CCDFGS03,Chi04,BACS05}.
Not only does this provide renewed motivation to search for quantum walk algorithms, but the specific construction may suggest new algorithmic approaches.  Second, this construction may provide tools for quantum complexity theory, as discussed further below.

A related result was shown by Feynman in one of the earliest papers on quantum computation, which constructs a Hamiltonian that can implement an arbitrary quantum computation \cite{Fey85}.  While Feynman's motivation was to give a physically reasonable model of a quantum mechanical computing device, his result can also be loosely interpreted as showing the universality of quantum walk.
In some sense, any time-independent Hamiltonian dynamics can be viewed as a quantum walk on a weighted, directed graph, with weights on opposite edges occurring in complex conjugate pairs to satisfy Hermiticity.  However, the interpretation of Hamiltonian dynamics as a quantum walk on a graph is clearest when the graph is sparse and the the edges are unweighted.
In Feynman's construction, the degree of the underlying graph grows with the size of the computation, and the edges must have weights.

This article presents an alternative Hamiltonian for universal quantum computation.
In this construction, the edges are unweighted, and the graph has maximum degree $3$ (which cannot be improved, since graphs of maximum degree $2$ clearly cannot be universal).
The main idea of the construction is to represent computational basis states by quantum wires, and to implement quantum gates by scattering off widgets attached to (and connecting) the wires.  For a circuit on $n$ quantum bits, the graph consists of $2^n$ wires, together with the attached widgets.  Each wire can be idealized as infinitely long, but in practice can be well approximated using only $\poly(n)$ vertices.  The computation begins at a single vertex on the far left side of one of the quantum wires, and the outcome of the computation corresponds to the wire on the far right side to which the state evolves under the quantum walk.

The remainder of this article is organized as follows.   In \sect{scatter}, we briefly review scattering theory on graphs, the main technical tool used in the paper.  In \sect{universal}, we describe the set of universal quantum gates used in the construction, and explain how to implement them by scattering processes at a certain fixed momentum.  In \sect{filter}, we show how undesirable momentum components can be removed by a filtering process, also achieved by scattering.  In \sect{compose}, we describe how the various elements of the construction behave under composition, and in \sect{bound}, we argue that bound states can be neglected.  In \sect{computer}, we summarize the construction of a universal quantum computer.  Finally, in \sect{discussion}, we conclude with a brief discussion of the results and some directions for future work.

\section{Scattering on graphs}
\label{sec:scatter}

First, consider an infinite line of vertices, each corresponding to a computational basis state $|x\>$ with $x \in \Z$, where vertex $x$ is connected to vertices $x \pm 1$.  The eigenstates of the adjacency matrix of this graph, parameterized by $k \in [-\pi,\pi)$, are the momentum states $|\tilde k\>$ with
\be
  \<x|\tilde k\> = e^{\ii k x}
\label{eq:kstate}
\ee
for each $x \in \Z$
(normalized so that
$
  \<\tilde k|\tilde k'\> = 2\pi \, \delta(k-k')
$),
with eigenvalues $2 \cos k$.

Now consider an arbitrary finite graph $G$, and create an infinite graph with adjacency matrix $H$ by attaching a semi-infinite line to each of $N$ of its vertices.  Label the basis states for vertices on the $j$th semi-infinite line as $|x,j\>$, with $x=0$ at the vertex in the original graph and $x=1,2,\ldots$ moving out along the line.  On each semi-infinite line, an eigenstate of the adjacency matrix must be a linear combination of states of the form \eq{kstate} with momenta $\pm k$, corresponding to an eigenvalue $2 \cos k$; or possibly of the same form but with $k=\ii\kappa$ or $k=\ii\kappa + \pi$ for some $\kappa > 0$, corresponding to an eigenvalue $2 \cosh \kappa$ or $-2\cosh\kappa$, respectively.  For each $j \in \{1,\ldots,N\}$ and each $k \in [-\pi,0]$,\footnote{Since the off-diagonal elements of the adjacency matrix are positive, whereas in a discrete approximation to the kinetic term $-{\d^2}/{\d{x}^2}$ they are negative, we use the convention that incoming states have \emph{negative} momentum.} there is an incoming scattering state of momentum $k$, denoted $|\tilde k, \incoming_j\>$, of the form
\ba
  \<x,j |\tilde k, \incoming_j\> &= e^{-\ii k x} + R_j(k) \, e^{\ii k x} \\
  \<x,j'|\tilde k, \incoming_j\> &= T_{j,j'}(k) \, e^{\ii k x}
  ,\quad
  j' \ne j
\ea
on the semi-infinite lines.  The reflection coefficient $R_j(k)$, the transmission coefficients $T_{j,j'}(k)$, and the form of $|\tilde k, \incoming_j\>$ on the vertices of $G$ are determined by the condition $H |\tilde k, \incoming_j\> = 2 \cos k |\tilde k, \incoming_j\>$.  For any fixed $k$, these coefficients can be found by solving $|G|$ linear equations.
Together with bound states $|\tilde\kappa,\bound^\pm\>$ (for certain discrete values\footnote{Since $\norm{H}$ is at most the maximum degree $d$ of the graph, $\kappa \le \cosh^{-1}(d/2)$.} of $\kappa > 0$ that can be obtained by solving $|G|-1$ linear equations and one transcendental equation) of the form
\be
  \<x,j |\tilde \kappa, \bound^\pm\> = B^\pm_j(\kappa) \, (\pm e^{-\kappa})^x,
\ee
the states $|\tilde k, \incoming_j\>$ form a complete, orthogonal set of eigenfunctions of $H$ that are useful for the analysis of scattering off the graph $G$.%

To calculate the propagator for scattering through $G$, we can expand the evolution in the basis of incoming scattering states and bound states.  For evolution from vertex $x$ on line $j$ to vertex $y$ on line $j' \ne j$, we have
\ba 
  &\<y,j'|e^{-\ii Ht}|x,j\> \nn
  &\quad=\sum_{\jbar=1}^N \intk{e^{-2\ii t\cos k}
    \<y,j'|\tilde k,\incoming_\jbar\>\<\tilde k,\incoming_\jbar|x,j\>\,} \nn
  &\qquad+\sum_{\kappa,\pm} e^{\mp 2\ii t\cosh\kappa}
    \<y,j'|\tilde\kappa,\bound^\pm\>\<\tilde\kappa,\bound^\pm|x,j\> \\
  &\quad\begin{aligned}= \intk{e^{-2\ii t\cos k}
    \Big[ & T_{j,j'} e^{\ii ky} (e^{\ii kx} + R_j^* e^{-\ii kx}) \\[-2pt]
        + & (e^{-\ii k y} + R_{j'} e^{\ii ky}) T_{j',j}^* e^{-\ii kx} \\
        + & \sum_{\jbar \notin \{j,j'\}} T_{\jbar,j'} e^{\ii ky}
                                  T_{\jbar,j}^* e^{-\ii kx}
        \Big]} \end{aligned} \nn
  &\qquad+\sum_{\kappa,\pm} e^{\mp 2\ii t\cosh\kappa}
    B^\pm_{j'}(\kappa) B^\pm_j(\kappa)^* (\pm e^{-\kappa})^{x+y} \\
  &\quad=\intk{e^{-2\ii t\cos k} \big( T_{j,j'} e^{\ii k(x+y)} 
                                    + T_{j',j}^* e^{-\ii k(x+y)} \big)} \nn
  &\qquad+\sum_{\kappa,\pm} e^{\mp 2\ii t\cosh\kappa}
    B^\pm_{j'}(\kappa) B^\pm_j(\kappa)^* (\pm e^{-\kappa})^{x+y}
,
\label{eq:evolution}
\ea
where $\dbar{k} := \d{k}/2\pi$, and where in the last step we have used unitarity of the $S$-matrix.%
\footnote{Similarly to the incoming scattering states, we can define outgoing scattering states $|\tilde k, \outgoing_j\>$ of the form
\bas
  \<x,j |\tilde k, \outgoing_j\> &= e^{\ii k x} + R_j(k)^* \, e^{-\ii k x} \\
  \<x,j'|\tilde k, \outgoing_j\> &= T_{j,j'}(k)^* \, e^{-\ii k x}
  ,\quad
  j' \ne j
\eas
on the semi-infinite lines.  The incoming states are related to the outgoing states by a transformation known as the scattering matrix, or $S$-matrix, with $|\tilde k,\incoming_j\> = \sum_{j'} S_{j,j'}(k) |\tilde k,\outgoing_{j'}\>$.  It can be shown that the $S$-matrix is unitary, and has the form $S_{j,j}(k)=R_j(k)$ and $S_{j,j'}(k)=T_{j,j'}(k)$ for $j \ne j'$.  In particular, the orthogonality of columns $j$ and $j'$ of $S$ implies that $T_{j,j'} R_j^* + R_{j'} R_{j',j}^* + \sum_{\jbar \notin \{j,j'\}} T_{\jbar,j'} T_{\jbar,j}^* = 0$, giving the cancellation in \eq{evolution}.
}
(We often suppress the dependence of the transmission and reflection coefficients on momentum for notational convenience.)

We will argue later that the contribution from the bound states can be neglected.  According to the method of stationary phase
(see for example \cite[Sec.\ II.3]{Won01}),
the integral over $k$ is dominated by those values where the derivative of the phase of the integrand vanishes.  The second term in the integrand can be shown not to have any stationary points, and the phase of the first term is given by
$
  k(x+y)+\arg T_{j,j'}(k)-2t\cos k
$,
which is stationary for
\be
  x+y+\ell_{j,j'}(k)=v(k)t,
\label{eq:stationary}
\ee
where
\be
  v(k) := \frac{\d}{\d{k}} 2 \cos k = -2\sin k
\label{eq:groupvel}
\ee
is the \emph{group velocity} at momentum $k$, and
\be
  \ell_{j,j'}(k) := \frac{\d}{\d{k}} \arg T_{j,j'}(k)
\label{eq:efflength}
\ee
is the \emph{effective length} of the path through $G$ from line $j$ to line $j'$.\footnote{If the graph $G$ is simply a line of $\ell$ edges, then the transmission coefficient is $T(k) = e^{\ii k \ell}$, and the effective length is precisely $\ell$.  In general, however, the effective length is momentum-dependent, i.e., the propagation is dispersive.}
Then for large $x+y$ we have \cite[Eq.\ 3.2]{Won01}
\be
  |\<y,j'|e^{-\ii H t}|x,j\>|
  \sim
  \frac{|T_{j,j'}(k^\star)|}{\sqrt{2\pi|c(k^\star)|}},
\label{eq:stationaryphase}
\ee
where $k=k^\star$ satisfies \eq{stationary}, and
\be
  c(k):=2t\cos k +\frac{\d^2}{\d{k}^2}\arg T_{j,j'}(k).
\label{eq:curvature}
\ee

While semi-infinite lines are convenient for the purpose of analysis, they can be replaced by long but finite lines to give a construction based on a finite graph (cf.\ \cite{FG98}).  This replacement does not significantly change the dynamics since the quantum walk on a line has a maximum propagation speed.  To see this, note that in \eq{groupvel}, a maximum group velocity of $2$ is obtained at $k = -\pi/2$.
Alternatively, consider the propagator on an infinite line with adjacency matrix $H$:
\ba
  \<y|e^{-\ii Ht}|x\>
  &= \int_{-\pi}^\pi \!\! e^{\ii k(y-x)-2\ii t \cos k} \, \dbar{k} \\
  &= (-\ii)^{y-x} J_{y-x}(2t)
\label{eq:lineprop}
,
\ea
where $J_\nu(t)$ is a Bessel function of order $\nu$.  Since $J_\nu(t)$ decays exponentially in $\nu$ when $\nu = t(1+\epsilon)$ for any fixed $\epsilon > 0$, \eq{lineprop} describes a wavefront moving with speed $2$.  Thus, provided the lengths of all the attached lines are large compared to twice the total evolution time, the effect of truncating the lines is negligible.

\section{Universal gate set}
\label{sec:universal}

We now show how to implement a universal set of quantum gates by scattering on graphs.
We use a universal gate set consisting of the controlled-not gate together with two single-qubit gates that generate a dense subset of $\SU(2)$.

\begin{figure}
\begin{picture}(89,40)(-15,0)
\put(-15,30){\makebox(10,10){$\alab$}}
\multiput(24, 5)(0,10){4}{\circle{2}}
\multiput(25,25)(0,10){2}{\line(1,0){20}}
\put(0,30){\makebox(20,10){\footnotesize$|00_\rmin\>$}}
\put(0,20){\makebox(20,10){\footnotesize$|01_\rmin\>$}}
\put(0,10){\makebox(20,10){\footnotesize$|10_\rmin\>$}}
\put(0, 0){\makebox(20,10){\footnotesize$|11_\rmin\>$}}
\put(25,15){\qbezier(0,0)(5,0)(10,-5)\qbezier(10,-5)(15,-10)(20,-10)}
\put(25, 5){\qbezier(0,0)(5,0)(9, 4)\qbezier(11,6)(15,10)(20,10)}
\multiput(46, 5)(0,10){4}{\circle{2}}
\put(50,30){\makebox(25,10){\footnotesize$|00_\rmout\>$}}
\put(50,20){\makebox(25,10){\footnotesize$|01_\rmout\>$}}
\put(50,10){\makebox(25,10){\footnotesize$|10_\rmout\>$}}
\put(50, 0){\makebox(25,10){\footnotesize$|11_\rmout\>$}}
\end{picture}
\hspace{2.5ex}
\begin{picture}(55,30)(0,-10)
\put(-1,20){\makebox(10,10){$\blab$}}
\put(0,0){\makebox(10,10){\footnotesize$|\rmin\>$}}
\put(14,5){\circle{2}}
\put(15,5){\line(1,0){20}}
\put(25,5){\circle*{2}}
\put(25,15){\circle*{2}}
\put(18,22){\circle*{2}}
\put(32,22){\circle*{2}}
\put(25,29){\circle*{2}}
\put(25,5){\line(0,1){10}}
\put(25,15){\line(1,1){7}}
\put(25,15){\line(-1,1){7}}
\put(25,29){\line(1,-1){7}}
\put(25,29){\line(-1,-1){7}}
\put(36,5){\circle{2}}
\put(40,0){\makebox(15,10){\footnotesize$|\rmout\>$}}
\end{picture} \bigskip\\
\begin{picture}(90,30)(-15,-20)
\put(-15,20){\makebox(10,10){$\clab$}}
\put(0,20){\makebox(15,10){\footnotesize$|0_\rmin\>$}}
\put(0, 0){\makebox(15,10){\footnotesize$|1_\rmin\>$}}
\multiput(19,5)(0,20){2}{\circle{2}}
\multiput(20,5)(0,20){2}{\line(1,0){30}}
\multiput(30,5)(0,10){3}{\circle*{2}}
\multiput(40,5)(0,10){3}{\circle*{2}}
\multiput(30,5)(10,0){2}{\line(0,1){20}}
\multiput(51,5)(0,20){2}{\circle{2}}
\put(55,20){\makebox(20,10){\footnotesize$|0_\rmout\>$}}
\put(55, 0){\makebox(20,10){\footnotesize$|1_\rmout\>$}}
\end{picture}
\hspace{2.5ex}
\begin{picture}(55,50)
\put(-1,40){\makebox(10,10){$\dlab$}}
\put(0,0){\makebox(10,10){\footnotesize$|\rmin\>$}}
\put(14,5){\circle{2}}
\put(15,5){\line(1,0){20}}
\multiput(25,5)(0,10){4}{\circle*{2}}
\put(25,5){\line(0,1){35}}
\multiput(25,43)(0,2){3}{\circle*{.75}}
\put(25,15){\line(1,0){20}}
\multiput(35,15)(10,0){2}{\circle*{2}}
\put(35,15){\line(0,1){10}}
\put(35,25){\circle*{2}}
\put(36,5){\circle{2}}
\put(40,0){\makebox(15,10){\footnotesize$|\rmout\>$}}
\end{picture}
\hspace{2.5ex}
\begin{picture}(55,36)(0,-14)
\put(-1,26){\makebox(10,10){$\elab$}}
\put(0,0){\makebox(10,10){\footnotesize$|\rmin\>$}}
\put(14,5){\circle{2}}
\put(15,5){\line(1,0){20}}
\multiput(25,5)(0,10){4}{\circle*{2}}
\put(25,5){\line(0,1){30}}
\put(32,20){\circle*{2}}
\put(25,15){\line(7,5){7}}
\put(25,25){\line(7,-5){7}}
\put(36,5){\circle{2}}
\put(40,0){\makebox(15,10){\footnotesize$|\rmout\>$}}
\end{picture}
\caption{Widgets used to construct a universal quantum computer.  Open circles indicate vertices where previous or successive widgets can be attached.
$\alab$ Controlled-note gate.
$\blab$ Phase shift.
$\clab$ Basis-changing gate.
$\dlab$ Momentum filter.
$\elab$ Momentum separator.}
\label{fig:widgets}
\end{figure}
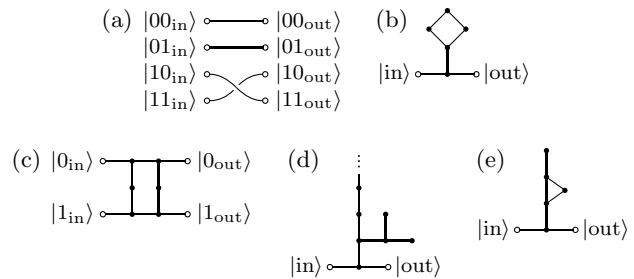

The controlled-not gate is trivial to implement.  This two-qubit gate exchanges the computational basis states $|10\>$ and $|11\>$, while leaving the other two states unchanged.  This transformation can be effected by simply exchanging the appropriate wires, using the widget shown in \fig{widgets}$\alab$.  An incoming wave of any momentum $k$ is transmitted perfectly through this widget,  accumulating a phase of $e^{\ii k}$.

\begin{figure}
\includegraphics[width=\columnwidth]{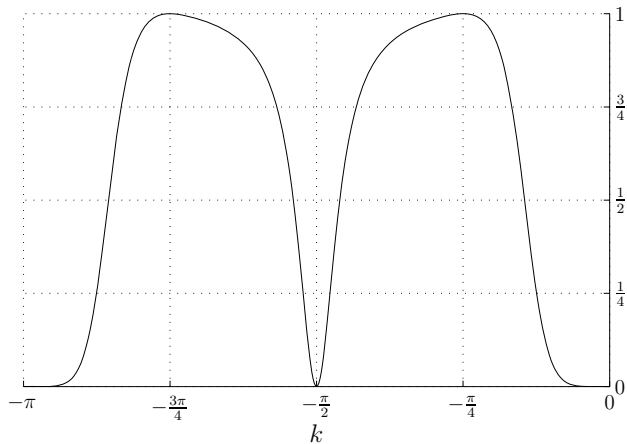}
\caption{Transmission probability for the phase shift widget (\fig{widgets}$\blab$).}
\label{fig:phaser}
\end{figure}

To implement a phase gate, we would like to apply some nontrivial phase to the $|1\>$ wire, while leaving the $|0\>$ wire unchanged.  This can be accomplished by inserting the widget shown in \fig{widgets}$\blab$ into the $|1\>$ wire.  To understand this widget, consider attaching semi-infinite lines to its terminals, and calculate the transmission coefficient for a wave of momentum $k$ incident on the input terminal.  We find
\be
  T_{\rmin,\rmout}^\blab = \frac{8}{8+\ii \cos 2k \csc^3 k \sec k},
\label{eq:btrans}
\ee
whose magnitude squared is plotted in \fig{phaser}.  In particular, this widget has perfect transmission at $k=-\pi/4$, where $T^\blab(-\pi/4)=1$ and $\ell^\blab(-\pi/4)=1$.  Relative to the effect of a straight wire of length $1$, the widget effectively introduces a phase of $e^{\ii\pi/4}$ at this momentum.  Combining the widget on the $|1\>$ wire with a straight wire for the $|0\>$ state, we see that for momenta near $-\pi/4$, the widget implements the phase gate
\be
  U_\rmb :=
  \begin{pmatrix}
    1 & 0 \\
    0 & e^{\ii\pi/4}
  \end{pmatrix}
\label{eq:phasegate}.
\ee
Note that momenta far from $-\pi/4$ (and $-3\pi/4$) will not only be transmitted with a different phase, but will also include a substantial reflected component.  However, we will see that the computation can be performed entirely with wave packets consisting of momenta near $-\pi/4$.

\begin{figure}
\includegraphics[width=\columnwidth]{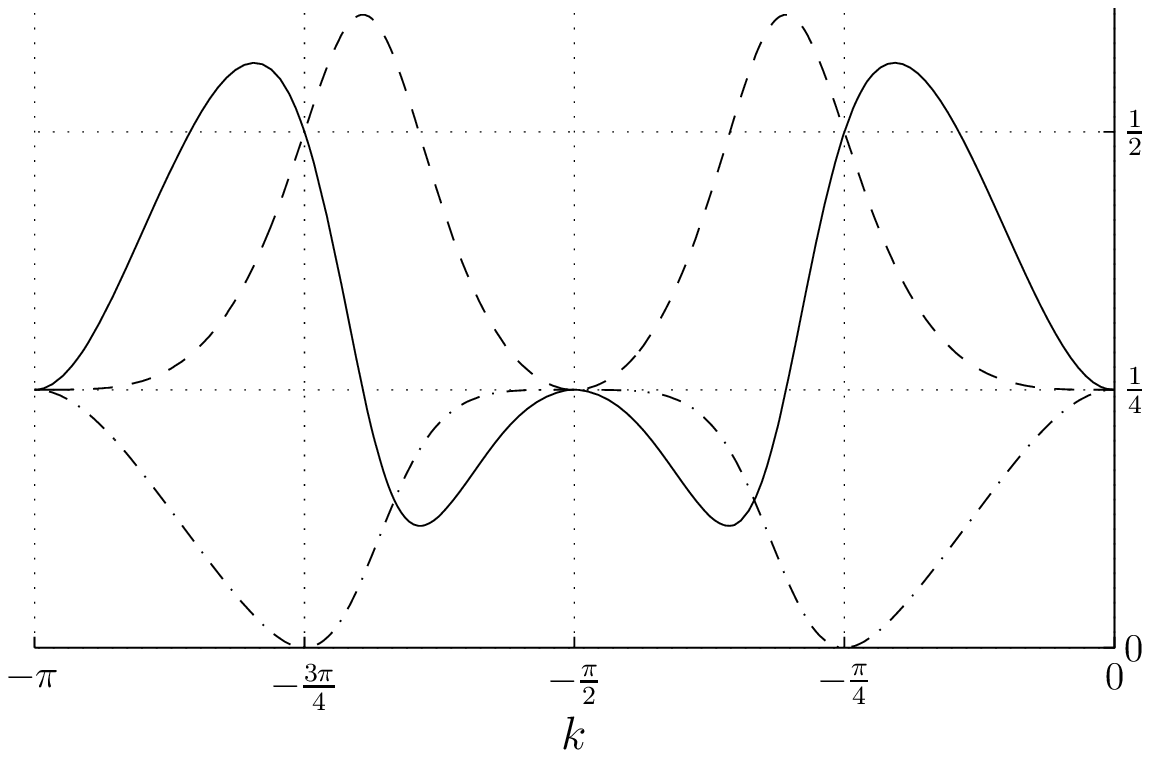}
\caption{Transmission probabilities for the basis-changing gate widget (\fig{widgets}$\clab$) with input at $|0_\rmin\>$ and outputs at $|0_\rmout\>$ (solid line), $|1_\rmout\>$ (dashed line), and $|1_\rmin\>$ (dot-dashed line).}
\label{fig:bcgate}
\end{figure}

To implement a basis-changing single-qubit gate, we must design a widget that includes interactions between different quantum wires.
Such a widget is shown in \fig{widgets}$\clab$.  To characterize this widget, we calculate the reflection and transmission coefficients for a wave of momentum $k$ incident on one terminal (say, the one labeled $|0_\rmin\>$ in \fig{widgets}$\clab$; the others are related by symmetry).  We find
\ba
  T_{0_\rmin,0_\rmout}^\clab &= \frac{e^{\ii k}(\cos k + \ii \sin 3k)}
                                     {2\cos k + \ii(\sin 3k - \sin k)} \\
  T_{0_\rmin,1_\rmout}^\clab &= -\frac{1}{2\cos k + \ii (\sin 3k - \sin k)} \\
  R_{0_\rmin}^\clab =
  T_{0_\rmin,1_\rmin}^\clab  &= -\frac{e^{\ii k} \cos 2k}
                                      {2\cos k + \ii (\sin 3k - \sin k)}.
\ea
The corresponding transmission probabilities are shown in \fig{bcgate}.
At $k=-\pi/4$, the input amplitude is transformed into an equal superposition of output amplitudes, with no amplitude reflected back to the input channels.  The effective lengths for forward transmission are $\ell^\clab_{0_\rmin,0_\rmout}(-\pi/4)=\ell^\clab_{0_\rmin,1_\rmout}(-\pi/4)=2$, so that the widget effectively lengthens the wires involved by two units.  Considering the phases of the transmission coefficients, we see that transmission through the widget effectively performs the unitary transformation 
\be
  U_\rmc :=
  -\frac{1}{\sqrt2}
  \begin{pmatrix}\ii & 1 \\ 1 & \ii\end{pmatrix}.
\label{eq:bcgate}
\ee
It is straightforward to show that this gate, together with the phase gate \eq{phasegate}, generate a dense subset of $\SU(2)$---for example, because $U_\rmb^2 U_\rmc U_\rmb^2$ is the Hadamard gate (up to a global phase) \cite{BMPRV99}.

\begin{figure}
\begin{picture}(195,86)(0,0)
\put(0,0){\makebox(20,10){\footnotesize$|11_\rmin\>$}}
\put(24,5){\circle{2}}
\put(25,5){\line(1,0){20}}
\put(35,5){\circle*{2}}
\put(35,15){\circle*{2}}
\put(28,22){\circle*{2}}
\put(42,22){\circle*{2}}
\put(35,29){\circle*{2}}
\put(35,5){\line(0,1){10}}
\put(35,15){\line(1,1){7}}
\put(35,15){\line(-1,1){7}}
\put(35,29){\line(1,-1){7}}
\put(35,29){\line(-1,-1){7}}
\put(45,5){\line(1,0){20}}
\put(55,5){\circle*{2}}
\put(55,15){\circle*{2}}
\put(48,22){\circle*{2}}
\put(62,22){\circle*{2}}
\put(55,29){\circle*{2}}
\put(55,5){\line(0,1){10}}
\put(55,15){\line(1,1){7}}
\put(55,15){\line(-1,1){7}}
\put(55,29){\line(1,-1){7}}
\put(55,29){\line(-1,-1){7}}
\put(85,5){\line(1,0){20}}
\put(95,5){\circle*{2}}
\put(95,15){\circle*{2}}
\put(88,22){\circle*{2}}
\put(102,22){\circle*{2}}
\put(95,29){\circle*{2}}
\put(95,5){\line(0,1){10}}
\put(95,15){\line(1,1){7}}
\put(95,15){\line(-1,1){7}}
\put(95,29){\line(1,-1){7}}
\put(95,29){\line(-1,-1){7}}
\put(105,5){\line(1,0){20}}
\put(115,5){\circle*{2}}
\put(115,15){\circle*{2}}
\put(108,22){\circle*{2}}
\put(122,22){\circle*{2}}
\put(115,29){\circle*{2}}
\put(115,5){\line(0,1){10}}
\put(115,15){\line(1,1){7}}
\put(115,15){\line(-1,1){7}}
\put(115,29){\line(1,-1){7}}
\put(115,29){\line(-1,-1){7}}
\put(0,32){\makebox(20,10){\footnotesize$|10_\rmin\>$}}
\put(24,37){\circle{2}}
\put(25,37){\line(1,0){20}}
\put(32,37){\circle*{2}}
\put(38,37){\circle*{2}}
\put(45,37){\line(1,0){20}}
\put(52,37){\circle*{2}}
\put(58,37){\circle*{2}}
\put(85,37){\line(1,0){20}}
\put(92,37){\circle*{2}}
\put(98,37){\circle*{2}}
\put(105,37){\line(1,0){20}}
\put(112,37){\circle*{2}}
\put(118,37){\circle*{2}}
\put(0,44){\makebox(20,10){\footnotesize$|01_\rmin\>$}}
\put(24,49){\circle{2}}
\put(25,49){\line(1,0){20}}
\put(35,49){\circle*{2}}
\put(35,59){\circle*{2}}
\put(28,66){\circle*{2}}
\put(42,66){\circle*{2}}
\put(35,73){\circle*{2}}
\put(35,49){\line(0,1){10}}
\put(35,59){\line(1,1){7}}
\put(35,59){\line(-1,1){7}}
\put(35,73){\line(1,-1){7}}
\put(35,73){\line(-1,-1){7}}
\put(45,49){\line(1,0){20}}
\put(55,49){\circle*{2}}
\put(55,59){\circle*{2}}
\put(48,66){\circle*{2}}
\put(62,66){\circle*{2}}
\put(55,73){\circle*{2}}
\put(55,49){\line(0,1){10}}
\put(55,59){\line(1,1){7}}
\put(55,59){\line(-1,1){7}}
\put(55,73){\line(1,-1){7}}
\put(55,73){\line(-1,-1){7}}
\put(85,49){\line(1,0){20}}
\put(95,49){\circle*{2}}
\put(95,59){\circle*{2}}
\put(88,66){\circle*{2}}
\put(102,66){\circle*{2}}
\put(95,73){\circle*{2}}
\put(95,49){\line(0,1){10}}
\put(95,59){\line(1,1){7}}
\put(95,59){\line(-1,1){7}}
\put(95,73){\line(1,-1){7}}
\put(95,73){\line(-1,-1){7}}
\put(105,49){\line(1,0){20}}
\put(115,49){\circle*{2}}
\put(115,59){\circle*{2}}
\put(108,66){\circle*{2}}
\put(122,66){\circle*{2}}
\put(115,73){\circle*{2}}
\put(115,49){\line(0,1){10}}
\put(115,59){\line(1,1){7}}
\put(115,59){\line(-1,1){7}}
\put(115,73){\line(1,-1){7}}
\put(115,73){\line(-1,-1){7}}
\put(0,76){\makebox(20,10){\footnotesize$|00_\rmin\>$}}
\put(24,81){\circle{2}}
\put(25,81){\line(1,0){20}}
\put(32,81){\circle*{2}}
\put(38,81){\circle*{2}}
\put(45,81){\line(1,0){20}}
\put(52,81){\circle*{2}}
\put(58,81){\circle*{2}}
\put(85,81){\line(1,0){20}}
\put(92,81){\circle*{2}}
\put(98,81){\circle*{2}}
\put(105,81){\line(1,0){20}}
\put(112,81){\circle*{2}}
\put(118,81){\circle*{2}}
\put(70,5){\line(0,1){32}}
\put(80,5){\line(0,1){32}}
\put(65,5){\line(1,0){20}}
\put(65,37){\line(1,0){20}}
\put(70,5){\circle*{2}}
\put(70,37){\circle*{2}}
\put(80,5){\circle*{2}}
\put(80,37){\circle*{2}}
\put(70,15){\circle*{2}}
\put(70,26){\circle*{2}}
\put(80,15){\circle*{2}}
\put(80,26){\circle*{2}}
\put(70,49){\line(0,1){32}}
\put(80,49){\line(0,1){32}}
\put(65,49){\line(1,0){20}}
\put(65,81){\line(1,0){20}}
\put(70,49){\circle*{2}}
\put(70,81){\circle*{2}}
\put(80,49){\circle*{2}}
\put(80,81){\circle*{2}}
\put(70,59){\circle*{2}}
\put(70,70){\circle*{2}}
\put(80,59){\circle*{2}}
\put(80,70){\circle*{2}}
\put(125,5){\qbezier(0,0)(10,0)(19.5,21)
            \qbezier(20.5,23)(24,30)(25,31)
            \qbezier(26,33)(32,44)(40,44)}
\put(125,49){\qbezier(0,0)(7,0)(14.5,-11)
             \qbezier(15.5,-13)(18,-17)(20,-22)
             \qbezier(20,-22)(30,-44)(40,-44)}
\put(125,37){\line(1,0){40}}
\put(125,81){\line(1,0){40}}
\put(166,5){\circle{2}}
\put(166,37){\circle{2}}
\put(166,49){\circle{2}}
\put(166,81){\circle{2}}
\put(170,0){\makebox(25,10){\footnotesize$|11_\rmout\>$}}
\put(170,32){\makebox(25,10){\footnotesize$|10_\rmout\>$}}
\put(170,44){\makebox(25,10){\footnotesize$|01_\rmout\>$}}
\put(170,76){\makebox(25,10){\footnotesize$|00_\rmout\>$}}
\end{picture}
\caption{Graph implementing a Hadamard gate on the second qubit followed by a controlled-not gate with the second qubit as the control.}
\label{fig:entangler}
\end{figure}
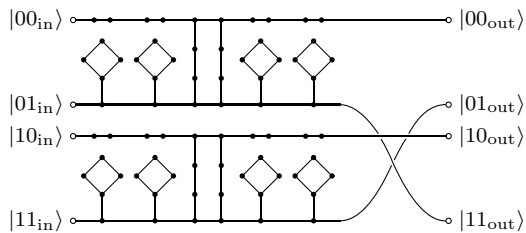

So far, we have only described how these gates act on one or two qubits at a time, but it is straightforward to embed them in a graph representing a computation on $n$ qubits.  For the controlled-not gate, we simply include its widget $2^{n-2}$ times, once for every possible setting of the $n-2$ qubits not involved in the gate.  Similarly, for the single-qubit gates, we include their widgets $2^{n-1}$ times.  As an example, \fig{entangler} shows the graph corresponding to a simple two-qubit quantum circuit.  Notice that, although the graph corresponding to an $n$-qubit circuit is exponentially large in $n$ (as it must be to represent an exponential number of basis states), it has a succinct description in terms of the original circuit being simulated.

Using only the three gate widgets $\alab$, $\blab$, and $\clab$, we can already construct a universal quantum computer, provided the input state is chosen appropriately.
Since there is no reflection at $k=-\pi/4$, the transmission coefficients at this momentum compose multiplicatively, so the concatenation of gate widgets can describe an arbitrary quantum circuit.
If the input state is prepared in a narrow wave packet consisting only of momenta close to $k=-\pi/4$, the propagation of this wave packet through the widgets implements that circuit.
However, we will see next that it is possible to use a much simpler starting state, corresponding to one particular vertex of the graph.

\section{Momentum filtering}
\label{sec:filter}

To construct a Hamiltonian that works with a simple starting state, we design a filter that only allows momenta near $k=-\pi/4$ to pass.  The basic building block of this filter is shown in \fig{widgets}$\dlab$.  Unlike the widgets for implementing gates, this widget includes a semi-infinite line, which allows undesired momentum components to be carried away.

\begin{figure}
\includegraphics[width=\columnwidth]{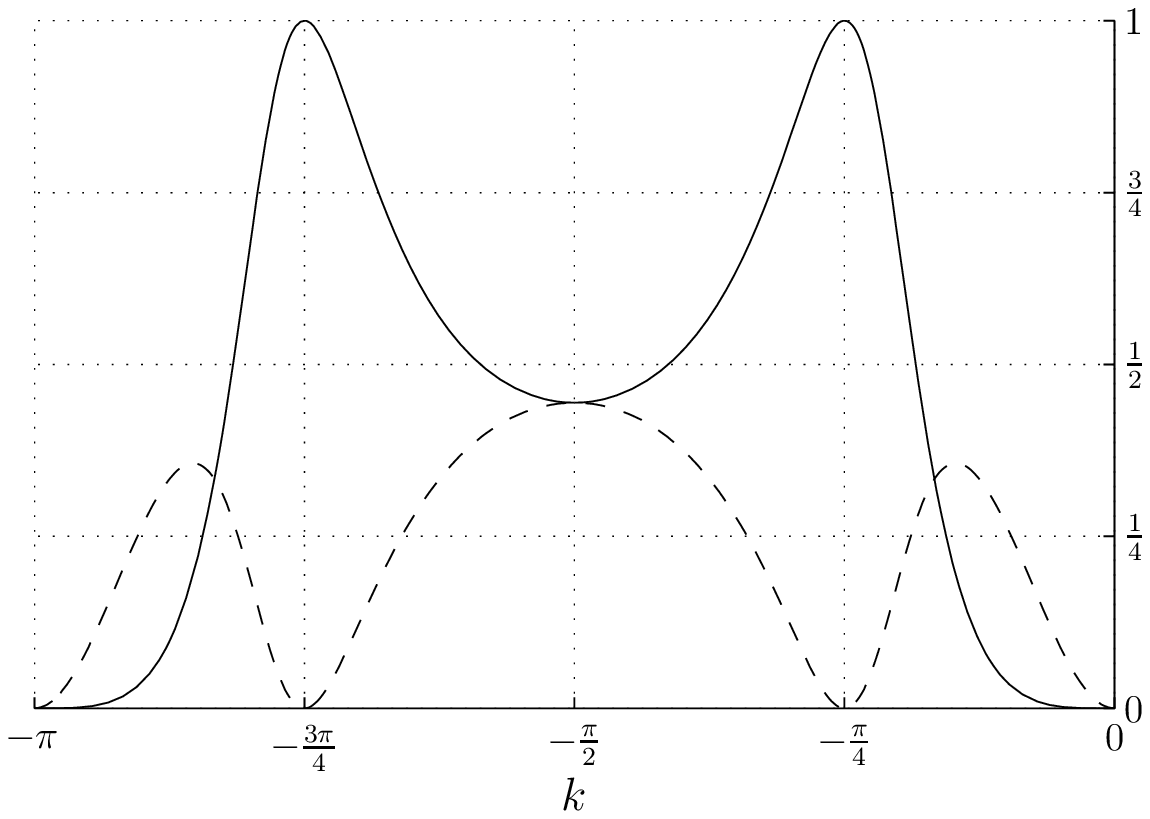}
\caption{
Transmission probabilities for the filter widget (\fig{widgets}$\dlab$) with input at $|\rmin\>$, and outputs at $|\rmout\>$ (solid line) and the semi-infinite line exiting upward (dashed line).}
\label{fig:filter}
\end{figure}

The transmission probabilities for this widget are shown in \fig{filter}.  Note that at $k=-\pi/4$ and $k=-3\pi/4$, all amplitude is transmitted forward, whereas at other momenta, some of the amplitude is transmitted upward and some is reflected. 

To filter out all but an arbitrarily narrow range of momenta, we repeat this widget many times in series.  The scattering properties of such a graph can be analyzed using a transfer matrix technique described in \cite{FG98}.
Consider an infinite line of vertices with additional edges attached to vertices $1,\ldots,m_\rmd$, and an arbitrary graph (here, the filter widget) attached above each of those vertices.  Then the amplitudes of the scattering state $|\tilde k,\incoming_\rmin\>$ satisfy
\be
  \begin{pmatrix}
    \<x+1|\tilde k, \incoming_\rmin\> \\
    \<x|\tilde k, \incoming_\rmin\>
  \end{pmatrix}
  = M \!
  \begin{pmatrix}
    \<x|\tilde k, \incoming_\rmin\> \\
    \<x-1|\tilde k, \incoming_\rmin\>
  \end{pmatrix}
\ee
where\footnote{This matrix differs from that in \cite{FG98} since we are using the adjacency matrix rather than the Laplacian as the Hamiltonian.}
\be
  M =
  \begin{pmatrix}
    2\cos k - y(k) \,& -1 \\
    1 & 0
  \end{pmatrix}
\label{eq:transfermatrix}
\ee
with
\ba
  y(k) &:= 
  \frac{\<\text{vertex above wire}|\tilde k, \incoming_\rmin\>}
       {\<\text{vertex on wire}   |\tilde k, \incoming_\rmin\>}
.
\ea
The transformation from the input amplitude to the output amplitude of the chain of $m_\rmd$ filter widgets is described by the matrix $M^{m_\rmd}$.  One can show that for
\be
  M^{m_\rmd} =
  \begin{pmatrix}
    a & b \\
    c & d
  \end{pmatrix},
\ee
the transmission coefficient is
\be
  T_{\rmin,\rmout}^\dlab
  =
  \frac{2 \ii e^{-\ii k m_\rmd} \sin k}
       {-a e^{-\ii k}-b+c+d e^{\ii k}}.
\label{eq:transfertransmit}
\ee

\begin{figure}
\includegraphics[width=\columnwidth]{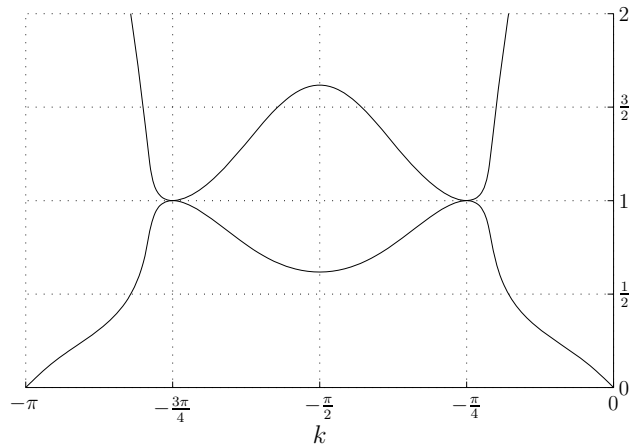}
\caption{Absolute values of eigenvalues of the transfer matrix.}
\label{fig:transfer}
\end{figure}

For the filter widget, a calculation shows\footnote{While $y(k)$ is real for any finite widget, the presence of a semi-infinite line allows it to be complex.}
\be
  y(k) = \ii e^{2 \ii k} \frac{\cos 2k}{\sin k}.
\ee
The absolute values of the eigenvalues of the corresponding transfer matrix \eq{transfermatrix} are shown in \fig{transfer}.  Except when $k$ is close to $-\pi/4$ or $-3\pi/4$, one of these eigenvalues is bounded above $1$; this eigenvalue gives the dominant contribution to \eq{transfertransmit}, and the transmission coefficient is exponentially small in $m_\rmd$.  However, $y(-\pi/4)=0$, so there is perfect transmission at $k=-\pi/4$.  Since $\ell_{\rmin,\rmout}^\dlab(-\pi/4)=2$, each instance of the widget effectively lengthens the wire by two units at this momentum.

Unfortunately, this filter transmits undesired momenta near $k=-3\pi/4$ as well as the desirable momenta near $k=-\pi/4$.  Generically, two distinct momentum components propagate at different speeds, and hence can be isolated temporally.  However, these particular momentum components have the same group velocity \eq{groupvel}.

\begin{figure}
\includegraphics[width=\columnwidth]{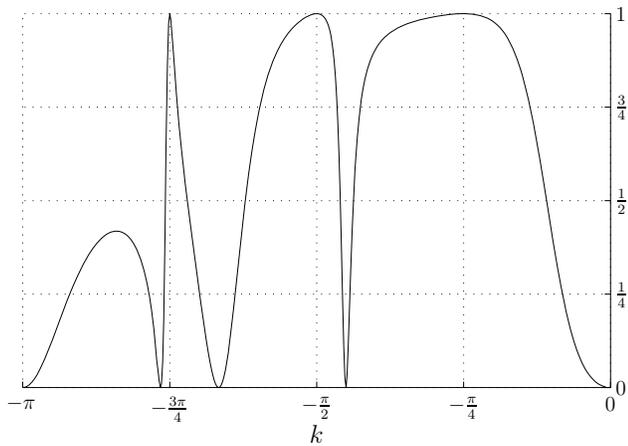}
\caption{Transmission probability for the momentum separator widget (\fig{widgets}$\elab$).}
\label{fig:separate}
\end{figure}

To isolate the two components, we can use the widget shown in \fig{widgets}$\elab$, which has a different effective length for $k=-\pi/4$ and $k=-3\pi/4$.\footnote{The symmetry of the transmission and reflection coefficients under $k \to -\pi-k$ for widgets $\alab$--$\dlab$ holds because those graphs are all bipartite.  Notice that widget $\elab$ breaks this symmetry.\label{foot:bipartite}}  The transmission coefficient for this widget is
\be
  T_{\rmin,\rmout}^\elab
  \!=\!
  \left[1+\frac{\ii(\cos k + \cos 3k)}
               {\sin k + 2\sin 2k + \sin 3k - \sin 5k}\right]^{-1}
\!\!,
\ee
as pictured in \fig{separate}.  There is perfect transmission at $k=-\pi/4$ and $-3\pi/4$ (as well as $k=-\pi/2$).  Furthermore, the derivative of the phase of the transmission coefficient gives $\ell^\elab_{\rmin,\rmout}(-\pi/4) = 4(3-2\sqrt2) \approx 0.686$ at $k=-\pi/4$, and $\ell^\elab_{\rmin,\rmout}(-3\pi/4) = 4(3+2\sqrt2) \approx 23.3$ at $k=-3\pi/4$.  Since the effective length of the widget is different for these two momenta, it serves to temporally separate them.

\section{Composing widgets}
\label{sec:compose}

It is straightforward to combine these widgets to simulate an arbitrary $m$-gate quantum circuit: simply connect the widgets in sequence corresponding to the gates in the original circuit.  For momentum $k=-\pi/4$, the transmission coefficients of the resulting graph exactly implement the desired circuit.  Furthermore, the effective lengths \eq{efflength} and curvatures \eq{curvature} at $k=-\pi/4$ simply add, so both are proportional to $m$ for the overall graph.  However, a propagating wave packet comprises a range of momenta, so we need to determine how close $k$ must be to $-\pi/4$ such that the transmission gives a good approximation to the desired circuit.  In turn, this determines how many filter widgets to include.

To understand how gate widgets behave under composition, it is helpful to view the scattering problem in terms of $2^n$ input channels scattering into $2^n$ output channels.  Define matrices $\cT,\cR,\bar\cT,\bar\cR$, describing forward transmission, reflection from forward to backward, backward transmission, and reflection from backward to forward, respectively, as
\ba
\label{eq:tmat}
  \cT_{j,j'}     &= T_{j_\rmin,j'_\rmout} &
  \cR_{j,j'}     &= \begin{cases}
                      R_{j_\rmin}          & j=j' \\
                      T_{j_\rmin,j'_\rmin} & j \ne j'
                    \end{cases} \\
  \bar\cT_{j,j'} &= T_{j_\rmout,j'_\rmin} &
  \bar\cR_{j,j'} &= \begin{cases}
                      R_{j_\rmout}           & j=j' \\
                      T_{j_\rmout,j'_\rmout} & j \ne j'
                    \end{cases}
\label{eq:rbarmat}
\ea
for $j,j' \in \{0,\ldots,2^n-1\}$.%
\footnote{For widgets symmetric under interchanging the roles of input and output, as is the case for all our widgets, $\bar\cR = \cR$ and $\bar\cT = \cT$.  However, this does not hold when composing distinct widgets.}
If we place two widgets in series, we can compute the transmission and reflection matrices for the composed widget as follows:
\ba
  \cT_{12}
  &= \cT_1 (1-\cR_2\bar\cR_1)^{-1} \cT_2 \label{eq:tcompose} \\
  \cR_{12}
  &= \cR_1 + \cT_1 (1-\cR_2\bar\cR_1)^{-1} \cR_2 \bar\cT_1 \label{eq:rcompose} \\
  \bar\cT_{12}
  &= \bar\cT_2 (1-\bar\cR_1\cR_2)^{-1} \bar\cT_1 \label{eq:tbarcompose} \\
  \bar\cR_{12}
   &= \bar\cR_2 + \bar\cT_2 (1-\bar\cR_1\cR_2)^{-1} \bar\cR_1 \cT_2 \label{eq:rbarcompose}
.
\ea
These expressions can be obtained by constructing the scattering states for the composed widget out of scattering states for the individual widgets.  For example, to construct $|\tilde k,\incoming_{j_\rmin}\>_{12}$, we begin by concatenating $|\tilde k,\incoming_{j_\rmin}\>_1$ (with its output wires removed) and $\sum_{j'} (T_1)_{j_\rmin,j'_\rmout}|\tilde k,\incoming_{j'_\rmin}\>_2$ (with its input wires removed).  This state fails to satisfy the eigenvalue condition only at the boundary between the widgets, so we attempt to correct this by including a reflection off widget $2$: we add $\sum_{j'} (T_1)_{j_\rmin,j'_\rmout} \big( (R_2)_{j'_\rmout} |\tilde k,\incoming_{j'_\rmout}\>_1 + \sum_{j'' \ne j'} (T_2)_{j'_\rmout,j''_\rmout} |\tilde k,\incoming_{j''_\rmout}\>_1\big)$ to the state of widget $1$.  Continuing to add terms corresponding to all the possible ways for waves to reflect between the two widgets, we obtain an infinite series that satisfies the eigenvalue condition even at the widget boundary.  Summing this geometric series, we obtain \eq{tcompose} and \eq{rcompose}, and a similar construction of the states $|\tilde k,\incoming_{j_\rmout}\>_{12}$ gives \eq{tbarcompose} and \eq{rbarcompose}.

Now we can see that the composition of two widgets, each with little reflection, also has little reflection.  In particular, suppose $\norm{\cR_1}, \norm{\bar\cR_1} \le \delta_1$ and $\norm{\cR_2}, \norm{\bar\cR_2} \le \delta_2$.  Then we have
\ba
  \norm{\cR_{12}}
  &\le \norm{\cR_1} + \Norm{(1-\bar\cR_1\cR_2)^{-1}} \norm{\cR_2} \\
  &\le \delta_1 + (1 + \delta_1 \delta_2)\delta_2,
\ea
as well as a similar bound for $\norm{\bar\cR_{12}}$.
Furthermore, forward transmission through the compound widget is nearly described by the product of the two forward transmission matrices, since
\ba
  \norm{\cT_{12} - \cT_1 \cT_2}
  &=   \Norm{\cT_1 \left(\frac{1}{1-\cR_2\bar\cR_1} - 1\right) \cT_2} \\
  &\le \Norm{\frac{1}{1-\cR_2\bar\cR_1} - 1} \\
  &= \Norm{\frac{\cR_2\bar\cR_1}{1-\cR_2\bar\cR_1}} \\
  &\le \delta_1\delta_2 (1+\delta_1\delta_2).
\ea

We can apply these bounds to study the transmission through $m$ gate widgets for momenta near $k=-\pi/4$.  In particular, suppose $|k+\pi/4| = O(1/m^2)$; then $\norm{\cR} = O(1/m^2)$ for each of the gate widgets.  Applying the bounds recursively, we find that the collection of all $m$ gate widgets has $\norm{\cR} = O(1/m)$ for such momenta; in other words, the transmission is nearly perfect.

To filter out all undesired momenta, suppose we precede the gate widgets by $m_\rmd=\log \Theta(m^2)$ filter widgets.  Then the output of the filter has exponentially small amplitude except for momentum components $k$ with $|k+\pi/4| = O(1/m^2)$ and $|k+3\pi/4| = O(1/m^2)$.

\section{Bound states}
\label{sec:bound}

It remains to show that bound states of the graph can be neglected.  Since bound states decay as $e^{-\kappa x}$, we can effectively ignore bound states with, say, $\kappa = \Omega(1/m^4)$ by starting the walk a distance $x=\Theta(m^4)$ from the first widget.  This choice increases the required running time of the simulation, but only by a polynomial factor.

Strictly speaking, the above argument leaves open the possibility that a large number of very weakly bound states could influence the scattering behavior.  However, since the bound states in question have $\kappa = O(1/m^4)$, they have the very similar energies $\pm 2 \cosh \kappa = \pm(2 + O(1/m^8))$.  Thus, for $t=O(m^4)$, the phases of the weakly bound states of type $+$ in \eq{evolution} are all the same up to $O(1/m^4)$, and similarly for the weakly bound states of type $-$.

To have nearly identical phases for both types of bound states, we can simply choose an evolution time $t$ for which the phase difference $|e^{+2\ii t}-e^{-2\ii t}|$ is zero.  For propagation precisely at momentum $k=-\pi/4$, corresponding to the group velocity $v(-\pi/4)=\sqrt 2$, we would ideally choose $t=(x+\ell)/\sqrt{2}$, where $\ell$ is the total effective length of all widgets at $k=-\pi/4$.  However, we can vary $t$ slightly without significantly affecting the contribution from the scattering states.  Specifically, we choose $t=\pi\floor{(x+\ell)/\sqrt{2}\pi}=((x+\ell)/\sqrt{2})(1+O(1/m^4))$.  This ensures that $e^{\mp 2\ii t\cosh\kappa} = e^{\ii\phi} + O(1/m^4)$, where the phase $\phi$ is independent of $\kappa$ and the choice of $\pm$.  In other words, all weakly bound states enter with approximately the same phase.  Moreover, the momentum for which the phase is stationary (i.e., satisfying \eq{stationary}) remains $k^\star = -\pi/4 + O(1/m^4)$, so the contribution from the scattering states is essentially unchanged.

Now observe that at $t=0$, the amplitude \eq{evolution} is $0$, and since the initial contribution from the scattering states is negligible, the initial contribution from the bound states is also negligible.  Since the relative phases between weakly bound states at a time $t=O(m^4)$ can only change by $O(1/m^4)$, the amplitude deviation caused by neglecting the bound states in \eq{evolution} is $O(1/m^4)$, which will turn out to be dominated by the contribution from the scattering states.

\section{Universal computer}
\label{sec:computer}

Overall, an arbitrary $m$-gate quantum circuit describing a unitary transformation $U$ on $n$ qubits can be implemented by quantum walk as follows.  On the input wire $0_\rmin$, we place $m_\rmd=\log\Theta(m^2)$ filter widgets, followed by a momentum separation widget.  We then add widgets for the $m$ gates in the circuit, each with $2^n$ inputs and $2^n$ outputs.  The $2^n$ input wires, $2^n$ output wires, and $m_\rmd$ filter wires are all truncated (as discussed at the end of \sect{scatter}) at length greater than $2t \approx \sqrt{2}(x+\ell)$ (say, length $2(x+\ell)$), where $\ell$ is the total effective length of the filter, momentum separation, and gate widgets.  Initially, the computer is prepared in the state $|x,0_\rmin\>$ corresponding to vertex $x$ on input line $0$, where $x=\Theta(m^4)$ as discussed above.  We evolve this state with the Hamiltonian given by the adjacency matrix of the graph for time $t=\pi\floor{(x+\ell)/\sqrt{2}\pi}=O(m^4)$.  Finally, we measure in the vertex basis.  If the outcome is on some output wire $s \in \{0,1\}^n$ (which happens with probability $\Omega(1/m^4)$), we output that $s$; otherwise we discard the result and start over.

Conditioned on obtaining a valid output, the statistics of this simulation closely reproduce those of the original quantum circuit.  According to \eq{evolution}, the amplitude to propagate to vertex $0$ on output line $s$ is well approximated by
\be
  \<0,s_\rmout|e^{-\ii H t}|x,0_\rmin\>
  \approx \intk{e^{\ii kx-2\ii t\cos k} T_{0_\rmin,s_\rmout}\!(k)\,},
\ee
which by \eq{stationaryphase} (and the construction of the graph) is approximately $\<s|U|0\> \times \Omega(1/m^2)$.

\section{Discussion}
\label{sec:discussion}

The construction described in this article shows that quantum walk is a universal computational primitive, in the sense that any quantum computation can be efficiently simulated by a quantum walk on a sparse, unweighted graph.  In addition to establishing the universality of quantum walk, it is possible that ideas from this construction could be applied elsewhere in quantum information processing.  We conclude by briefly mentioning some of these possibilities.

One potential application is the design of quantum algorithms based on scattering on graphs.  In such algorithms, we need not think of scattering as directly implementing quantum gates on one or two qubits, but rather, as a generic way of implementing transformations on quantum states.  An early proposal along these lines was made in \cite{FG98} in the context of decision trees, and an algorithm for evaluating balanced binary game trees using similar ideas was given in \cite{FGG07}.  Indeed, generalizations of this approach have led to quantum algorithms for evaluating broad classes of formulas \cite{ACRSZ07,RS07}.

Another possible application is to quantum complexity theory.  Feynman's original quantum computer Hamiltonian \cite{Fey85} has been used to construct QMA-complete \cite{KSV02} and BQP-complete \cite{WZ06,JW06} problems, in addition to showing the universality of adiabatic evolution \cite{ADKLLR04,KR03,KKR04}.  The quantum walk construction could potentially be applied to construct new complete problems or adiabatic quantum computers with desirable properties.

Finally, this construction might be useful in the development of new architectures for quantum computers.  Although we have not concerned ourselves with the physical representation of the computer, the graph inherits a kind of tensor product structure from the underlying circuit, so it might be possible to encode the system such that the Hamiltonian involves only local interactions between qubits.  Such a construction could potentially be used to design a quantum computer that does not require dynamic control \cite{Ben80,Fey85,Mar90,JW05,Jan07,VC07,Kay08,NW08,CL08}.

\acknowledgments

I thank Shengyu Zhang for discussions that helped to motivate this work, and Edward Farhi and Jeffrey Goldstone for comments on a preliminary version of the manuscript.  In particular, the observation in footnote~\ref{foot:bipartite} is due to Goldstone.
This work was done in part while the author was at the Institute for Quantum Information at the California Institute of Technology, where he received support from the National Science Foundation under grant no.\ PHY-0456720 and from the Army Research Office under grant no.\ W911NF-05-1-0294.  This work was also supported in part by NSERC and MITACS.


\newcommand{\noopsort}[1]{} \newcommand{\printfirst}[2]{#1}
  \newcommand{\singleletter}[1]{#1} \newcommand{\switchargs}[2]{#2#1}

\end{document}